\documentclass[9pt,twocolumn,twoside]{osajnl}
\journal{ol} % Choose journal (ao, aop, josaa, josab, ol, optica, pr)

\setboolean{shortarticle}{true}
\usepackage{amsmath,mathrsfs}
\usepackage{xcolor}
\usepackage{graphicx}% Include figure files
\usepackage{dcolumn}% Align table columns on decimal point
\usepackage{bm}% bold math
\usepackage{slashbox}
\title{Robust and Efficient Single-Pixel Image Classification with Nonlinear Optics}

\author[1,2,*]{Santosh Kumar}
\author[1,2,*]{Ting Bu}
\author[1,2]{He Zhang}
\author[1]{Irwin Huang}
\author[1,2,**]{Yuping Huang}

\affil[1]{Department of Physics, Stevens Institute of Technology, Hoboken, NJ, 07030, USA}
\affil[2]{Center for Quantum Science and Engineering, Stevens Institute of Technology, Hoboken, NJ, 07030, USA}
\affil[*]{These authors contributed equally}
\affil[**]{Corresponding author: yhuang5@stevens.edu}

\date{\today}% It is always \today, today,
\begin{abstract}
We present a hybrid image classifier by mode-selective image upconversion, single pixel photodetection, and deep learning, aiming at fast processing a large number of pixels. It utilizes partial Fourier transform to extract the signature features of images in both the original and Fourier domains, thereby significantly increasing the classification accuracy and robustness. Tested on the MNIST handwritten digit images, it boosts the accuracy from 81.25\% to 99.23\%, and achieves an 83\% accuracy for highly contaminated images whose signal-to-noise ratio is only -17 dB. Our approach could prove useful for fast lidar data processing, high resolution image recognition, occluded target identification, atmosphere monitoring, and so on.   

\end{abstract}

\setboolean{displaycopyright}{true}

\begin{document}

\maketitle

%\section{Introduction}
%\label{sec:intro}

Machine learning techniques based on deep neural networks (DNNs) have scored considerable success in image classification \cite{Rosso2016,Sahlol2020}, speech recognition \cite{deng2013new}, image generation \cite{imagenet17}, image reconstruction \cite{Wang2020}, and so on. They use a feed-forward multi-layer neural architecture to achieve high performance for complex operations \cite{SCHMIDHUBER201585,Goodfellow-et-al-2016}. Despite notable success for their digital implementations, it remains a challenge to adopt them directly for embedded systems due to the high memory-demand, limited scalability, and excessive energy budget \cite{Embedded_Systems20}. On the other hand, optical systems naturally offer extreme parallelism and multiplexing yet with little energy consumption and fast data-processing \cite{Chakraborty_PhysRevApplied,PhysRevX_ONN19,Feldmann2021}. Thus far, optical Fourier transformation, diffraction, interference, and filtering have been exploited for optical pattern recognition \cite{Jiao:19,Zhang:20} and restoration \cite{Photonics20}, phase retrieval \cite{PRL_photonphaseRet,Qian:20}, and optical information processing \cite{steinbrecher_quantum_2019}.
Recently, optically assisted image processing has become a vivid pursuit in the field of artificial intelligence \cite{Wetzstein2020}, computer vision \cite{LATORRECARMONA2019}, robotics \cite{Shin2020,Xueaaw6304}, medical science \cite{Sullivan2018}, and so on.

While most of these studies are based on linear optics, nonlinear optics is poised to enable even richer and more complex operations and lift our processing capability to yet another level. To this end, optical neural network with nonlinear activation functions was implemented for object identification and classification of ordered and disordered phases of Ising models \cite{zuo_all-optical_2019}. Spontaneous parametric down-conversion was utilized for quantum-correlated pattern recognition with spatially structured photons \cite{QuantumPR2019}. Artificial neural networks based on nonlinear optics have shown advantages in reconstructing the amplitude and phase profiles of ultrashort pulses \cite{ziv_deep_2020,zahavy_deep_2018} and neuromorphic computing \cite{PhysRevLett_Conti,ballarini2020p}. A super-Ising emulator was demonstrated with unprecedented four-body interaction using spatial light modulation and second-harmonic generation \cite{Kumar2020}. 

Recently, we have proposed and experimentally demonstrated a nonlinear-optics approach to pattern recognition with single-pixel imaging and a deep neural network \cite{bu2020single}.
It employs mode-selective image up-conversion to project a raw image with up to mega pixels onto a set of coherent spatial modes, whereby its signature features are extracted optically in a nonlinear manner. Our experimental results promise distinct applications in online classification of large-size images, fast lidar data analyses \cite{Rehain2020}, complex pattern recognition, and so on.

In this work, we continue to study the aforementioned hybrid machine learning technique, focusing on further improving the efficiency and effectiveness in the feature extraction. In our previous system \cite{bu2020single}, the mode-selective conversion is applied to the fully Fourier-transformed images, which samples the signature features in the Fourier domain, but overlooks some important identifying features in the original image domain. Here, we study a new approach where the upconversion is applied to the images undergoing only partial Fourier transform. While challenging to calculate partial Fourier transform numerically \cite{Bowman2018,park2021fast}, it is easily realized in the current system by displacing the nonlinear crystal from the focal point of the Fourier lens for the images. This simple adjustment allows both the original and Fourier features of the images to be sampled at the same time, leading to high classification accuracy using only a limited number of pump modes for the upconversion. Tested on the MNIST (Modified National Institute of Standards and Technology) handwritten digit images, a 40 mm crystal displacement significantly increases the accuracy from 81.25\% to 99.23\%. The same technique enables robust pattern recognition when the images are mixed with strong noise, where a high accuracy $\sim 83\%$ is achieved even when the signal-to-noise (SNR) is as low as -17 dB. All of our experimental results agree well with the simulated results, confirming this effect. Finally, we numerically compare two choices of the pump mode bases for the upconversion that can affect the feature extraction, showing better performance with the Laguerre Gaussian (LG) mode bases than the Hermite-Gaussian (HG) modes, likely due to a lower cross-talk for the former \cite{Restuccia:16}. Our results highlight a viable approach to image recognition through feature extraction in both the original and Fourier domains by mode-selective image upconversion, for high accuracy, high efficiency, and exceptional robustness.

\begin{figure}[ht]
\centering
\includegraphics[width=\linewidth]{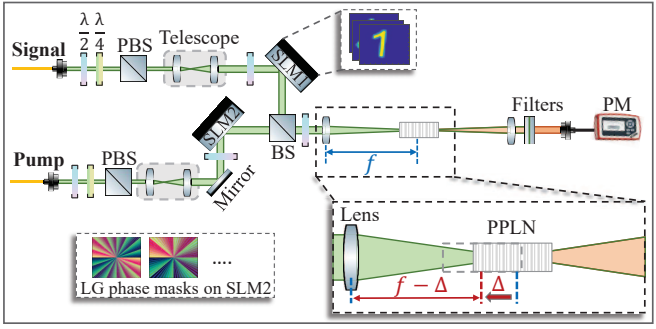}
\caption{Experimental setup for the feature extraction via frequency up-conversion.  
The PPLN crystal is moved away from the focal point to the focal lens by a distance $\Delta$, as shown in the lower right inset. SLM: Spatial Light Modulator, BS: Beamsplitter, WDM: Wavelength-division multiplexing, PPLN: Magnesium-doped Periodic Poled Lithium Niobate crystal; PM: power meter.}%Both beams are focused using a lens of focal length $f=20$cm.
\label{fig:setup}
\end{figure}

%\section{Experimental setup}
%\label{sec:experiments}

The experimental setup for the present nonlinear optical pattern-recognition scheme is outlined in Fig.~\ref{fig:setup}. The phase patterns of the Gaussian beams in signal and pump arms are modulated by two separate spatial light modulators (SLM). In the signal arm, SLM1 is used to upload phase patterns of input images while the SLM2 is applied to map phase patterns of spatial LG (or HG) modes on pump beam. The phase value is wrapped in the interval between 0 and $2\pi$ to express on the SLMs \cite{Santosh19,zhang_mode_2019}. The phase modulated two beams are then merged at a beam splitter (BS). A Fourier lens ($f$=200 mm) is used to focus the coupled beam in a temperature-stabilized periodic poled lithium niobate (PPLN) crystal for sum-frequency (SF) generation process. The power readings of the SF generation are extracted features from images, which are inputs to a deep neural network for digits recognition. Further details of our setup can be found in our previous work \cite{bu2020single}. The difference between the current setup and the previous one is the position of the PPLN. As illustrated in the lower right inset in Fig.~\ref{fig:setup}, $\Delta$ is the of the displacement of PPLN crystal towards the Fourier lens.

In simulation, for each pair of the signal beam $E_s$ and pump beam $E_p$, their corresponding SF field $E_f$ is solved under the slowly-varying-envelope approximation as: \begin{eqnarray}
2ik_{s}\partial_z E_{s}+(\partial_{x}^{2}+\partial_{y}^{2})E_{s}=-2\frac{\omega_{s}^{2}}{c^{2}}\chi^{(2)} E_{p}^{*}E_{f}e^{i\triangle k z},
\label{eqES}\\
%2
%\begin{equation}
2ik_{p}\partial_zE_{p}+(\partial_{x}^{2}+\partial_{y}^{2})E_{p}=-2\frac{\omega_{p}^{2}}{c^{2}}\chi^{(2)} E_{s}^{*}E_{f}e^{i\triangle k z},
\label{eqEP}\\ 
%\end{equation}
%3
%\begin{equation}
2ik_{f}\partial_zE_{f}+(\partial_{x}^{2}+\partial_{y}^{2})E_{f}=-2\frac{\omega_{f}^{2}}{c^{2}}\chi^{(2)} E_{p}E_{s}e^{-i\triangle k z}, \label{eqEF}
\end{eqnarray}
where $\omega_{s}$, $\omega_{p}$, and $\omega_{f}$ are the beam waists of signal, pump and SF light, respectively.  
$k_{s}={n_{s}\omega_{s}}/{c}$ , $k_{p}={n_{p}\omega_{p}}/{c}$ and $k_{f}={n_{f}\omega_{f}}/{c}$
are their wave numbers. $\Delta k=k_{s}+k_{p}-k_{f}-2\pi/\Lambda$ gives their phase mismatch, $\Lambda =19.36$ $\mu$m is the poling period of the PPLN. $z$ is the distance in propagation direction with $z=0$ being the focal point. $\chi^{(2)}$ is the second-order nonlinear susceptibility.
We can use the standard split-step Fourier and adaptive step size methods \cite{agrawal2007nonlinear} to numerically solve  Eqs.(\ref{eqES})-(\ref{eqEF}) with an initial location $z=-\frac{L}{2}-\Delta$, where $L=1$ $cm$ is the length of the PPLN.

\begin{figure}[htp]
\centering
\includegraphics[width=1\linewidth]{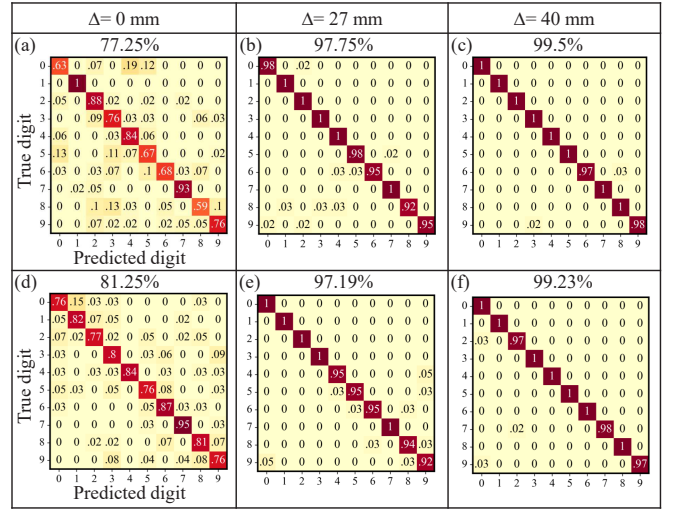}
\caption{Normalized confusion matrices for testing the performance of the hand-written digits with three different crystal positions. The top panels (a-c) are the simulated results using 40 LG pump modes with $\Delta = 0$, 27 mm and 40 mm, respectively. The bottom panels (d-f) are the corresponding experimental results. % (a) and (b) $\Delta = 0$, (b) and (e) for $\Delta =$ 27 mm, and (c) and (f) for $\Delta =$ 40 mm.
}
\label{fig:exp_conf} 
\end{figure}

%\section{Data Processing }
%\label{sec:data}
In this paper, one of the most popular MNIST handwritten digit database is selected as our benchmark, which is a 10-class database including handwritten images representing digits from "0" to "9" \cite{lecun2010}. 
A subset of the original database consisting of the first 200 handwritten images of each digit is applied as our database.
%In our database, the 200 handwritten images of each digit are used for training and benchmarking. 
These images in our database are shuffled and separated into training (1600 images) and testing (400 images) sets.
The resolution of these 28$\times$28 images are increased to 400$\times$400 pixels to match with the input signal beam size and SLM pitch.
% Meanwhile, the pump modes are chosen as a set of LG or HG spatial modes to drive the mode-selective upconversion for features extraction, which are the inputs to a deep neural network for digits recognition.
The SF power readings of each handwritten image are normalized between 0 and 1, after collecting all results for our training and testing data sets. This normalization is different from our previous work in \cite{bu2020single}, where the upconversion takes place in the Fourier plane and data are normalized within each LG mode. While the original and Fourier planes offer equivalent descriptions \cite{goodman2005}, the amount of information that can be extracted using a small number of pump modes is quite different. 
In \cite{bu2020single}, the upconversion was performed in the Fourier domain, which resulted in similar SF power distributions for the first few pump modes. The signature features are contained in the higher order modes where the SF power is relatively low. 
To distinguish those digits therefore requires ``amplifying'' the SF power difference in the higher order modes, which was realized through SF power normalization by mode.
While high performance is permissible, that method renders the system prone to noise and fluctuations. Also, after adding new training or testing data, the whole data set needs to be normalized again and new training process is warranted for the best performance.

\begin{figure}[htp]
\centering
\includegraphics[width=1\linewidth]{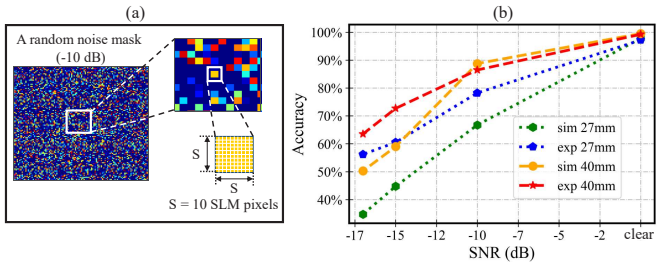}
\caption{(a) An example of random noise mask with $\textrm{SNR}=-10$ dB. The zoomed windows are representing the noise clusters with cluster size S equal to the length of 10 SLM pixels (i.e. one noise pixel size, $\textrm{S}=10$ SLM pixels or 104 $\mu$m). (b) Experimental and simulated accuracy vs SNR. Blue and green dotted curves are the experimental and simulated accuracy for $\Delta$ = 27 mm, respectively, and red and orange dashed curves are for $\Delta$ = 40 mm.}
\label{fig:acc_vs_snr} 
\end{figure}

In contrast, when the upconversion is applied to partially Fourier transformed images, the signature features of the images are extracted in both the original and Fourier domains. They are readily manifested in SF power for the first few LG modes of high readings, resulting in distinct power distribution over the 40 LG modes. This eliminates the need to amplify the SF power for higher-order modes, and allows normalization for each image. It makes the system more robust and efficient, while avoiding the need of repeated normalization within each mode over the entire data set when new data are added. 

The structure of the deep neural network is same as the one in our previous work in \cite{bu2020single}, which is a combination of one convolutional layer and five fully connected layers. In the training process, the  adaptive moment estimation (ADAM) gradient descent algorithm is selected to minimize the loss between predictions and their ground truth. The activation function softmax is applied to normalize the output values between 0 and 1 so that they can represent the probabilities of each class. Then the predicted class of each input image will be chosen as the one with the largest probability. Finally, the classification accuracy can be calculated as the fraction of correctly classified images.

%\section{results}
%\label{sec:results}

To optimally extract the spatial and Fourier features at the same time, we displace the PPLN crystal closer to the Fourier lens with its center away from the focal point by $\Delta$, as shown in Fig.\ref{fig:setup}. 
While this leads to efficient and robust classification, a large $\Delta$ de-focuses the light and suppresses the SF generation. 
In light of this trade-off, we choose  $\Delta=0$, 27 mm and 40 mm in this study. 
A larger $\Delta$ may be accommodated by using a higher pump power to compensate for the SF efficiency loss. 
Also, here we follow \cite{bu2020single} to use a set of 40 LG modes, with $l \in [-2, 2]$ and $p \in [0, 7]$. Figure \ref{fig:exp_conf} lists the normalized confusion matrices using the 40 LG pump modes for features extraction. Figure \ref{fig:exp_conf}(a-c) are the simulated results and (d-f) are the experimental results with $\Delta = 0,$ 27 mm and 40 mm, respectively. 
When the crystal is centered at the Fourier plane
with $\Delta$ = 0, the setup is the same with that in \cite{bu2020single}. 
In this case, the simulated accuracy is only 77.25\% and experimental accuracy is 81.25\%. 
This discrepancy is likely attributed to the measurement errors in the upconverted SF power readings for high-order pump projection modes. 
When the crystal is displaced by $\Delta = 27$ mm, both the simulated and experimental accuracy reached above 97\% as shown in Fig. \ref{fig:exp_conf}(b) and (e), respectively. This shows the advantage of feature extraction in both spatial and Fourier domains. Figure \ref{fig:exp_conf}(c) and (f) show that the accuracy can be further increased over 99\% with $\Delta = 40$ mm.

To test the robustness of our technique against noise, we next mix the images with uniformly distributed random noise. 
The noise level is quantified by the signal-to-noise ratio (SNR), defined as:
\begin{equation}
   \textrm{SNR}=10~\textrm{log}_{10}(\sigma _{s}^{2}/\sigma _{n}^{2}),
\label{eq:snr}
\end{equation}
where $\sigma _{s}^{2}$ and $\sigma _{n}^{2}$ are the variance of per-pixel phase values in the digit images and the added noise masks, respectively.
Figure \ref{fig:acc_vs_snr}(a) shows an example of one uniformly distributed noise mask with SNR = -10 dB and cluster size $\textrm{S}=104$ $\mu$m, where the cluster size is the number of pixels in one noise cluster (which is 10 SLM pixels in this case). The white noise masks are added to the original handwritten digit images and are uploaded on SLM1 as the inputs. The results in Fig. \ref{fig:acc_vs_snr}(b) show that, with a larger SNR, the accuracy is higher and the difference is smaller between the simulated and experimental results. Similar to those in Fig.~\ref{fig:exp_conf} for clear images, for the noise-contaminated images, both results reveal better performance with $\Delta$ = 40 mm than $\Delta$ = 27 mm. The advantage reaches 12.18\% when the SNR is -15 dB. This again validates our approach for efficient and robust information extraction. 
 
\begin{figure}[htp]
\centering
\includegraphics[width=0.9\linewidth]{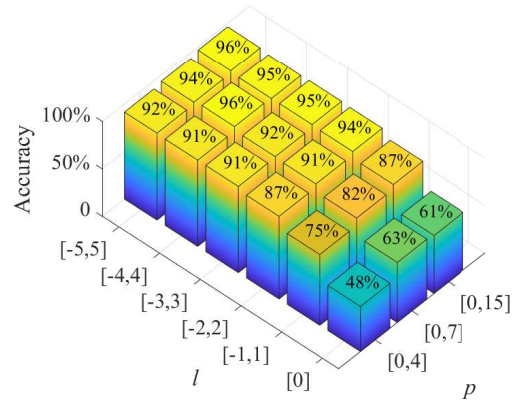}
\caption{The accuracy for different sets of LG modes with different azimuthal ($l$) and radial ($p$) indexes, with $\textrm{SNR}=-15$ dB and $\textrm{S}=52 $ $\mu$m. }
\label{fig:more_modes} 
\end{figure}
In addition to SNR, the cluster size S is also an import factor to the classification accuracy. When the noise feature size is comparable with the image's characteristic features, it is difficult to distinguish them in both original and Fourier domains. The simulated and experimental results with SNR=-15 dB and $\Delta$ = 40 mm in Fig.~\ref{fig:acc_vs_snr} (b) are 59\% and 72.68\%, respectively. However, when we reduce the cluster size by half ($\textrm{S}=52 $ $\mu$m), the accuracy increases to 91.25\% and 93.4\% in simulation and experiment, respectively.

Thus far, 40 LG modes \{LG$^p_l$\} are chosen for the pump, with $l \in [-2, 2]$ and $p \in [0, 7]$, to achieve the accuracy for clear images while aiming at high efficiency. In practice, one may further optimize the mode bases for a particular application. To illustrate this opportunity, Fig.~\ref{fig:more_modes} plots the simulated accuracy using different sets of LG mode basis with $\textrm{SNR}=-15$ dB and  $\textrm{S}=52 $ $\mu$m.
It shows that with more modes, the accuracy can be significantly higher when more than 40 LG modes are used. With 176 LG modes ($l \in [-5, 5]$ and $p \in [0, 15]$), the accuracy improved to 96\% from 91\%. On the other hand, when the number of LG modes is smaller than the 40 modes, the accuracy could drop quickly from 87\% with $l \in [-2, 2]$ and $p \in [0, 4]$ to 48\% with $l = 0$ and $p \in [0, 4]$). 

\begin{figure}[htp]
\centering
\includegraphics[width=1\linewidth]{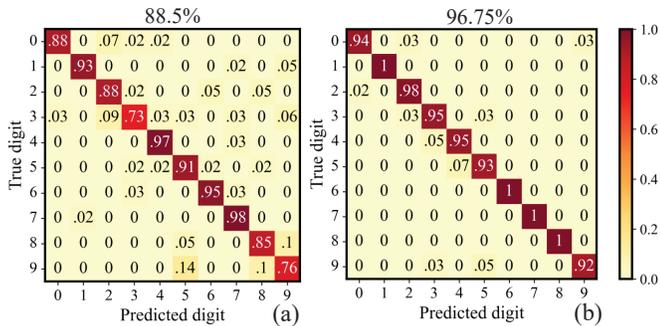}
\caption{Simulated normalized confusion matrices using 49 HG pump modes with (a) $\Delta$ = 27 mm and (b) $\Delta$ = 40 mm.}
\label{fig:sim} 
\end{figure}

Lastly, we compare the choices of LG and HG pump modes to drive the upconversion for image classification. HG modes are another popular class of orthogonal mode basis which, unlike LG modes of rotational symmetric intensity, exhibit rectangular symmetry in intensity. Figure \ref{fig:sim} shows the simulated results with the 49 lowest order HG modes, \{HG$_{mn}$\} with $m$ and $n \in [0, 7]$. The classification accuracy of the same data base is 88.5\% with $\Delta = 27$ mm and 96.75\% with $\Delta = 40$ $ mm$, which are all inferior to the simulated results with 40 LG modes in Fig. \ref{fig:exp_conf}(b) and (c). Therefore, LG modes are more efficient and suitable for the feature extractions of the handwritten digits than HG modes. Of course, in practice the optimum mode choice shall depend on the spatial characteristics of the target images. 

%\section{Conclusion}
In summary, we have demonstrated a new approach to efficient and robust image classification using a hybrid system integrating nonlinear optics and a deep neural network. It combines mode selective frequency up-conversion and partial Fourier transformation to sample the signature features in the original and Fourier-transformed images through a single nonlinear optical stage. The partial Fourier transform is conveniently implemented by moving the up-conversion crystal away from the focal point. It significantly improves the classification accuracy for both clear and noise-contaminated images. Our results tap on an inherent advantage of coherent optics for complex machine learning tasks: the ease and flexibility of linear algebra operations for high volume data. The robustness of this approach against strong noise invites its future applications in challenging tasks such as identifying occluded targets, wide-field surveillance, and remote sensing under low visibility.

%\section*{Acknowledgments}
\noindent\textbf{Disclosures.} The authors declare no conflicts of interest.

%\nocite{*}
\bibliography{ref}

 \bibliographyfullrefs{ref}

\end{document}